\documentclass{article}
\usepackage{spconf,amsmath,graphicx}
\usepackage{subfigure}
\usepackage{multirow}
\usepackage{array,url}
\usepackage{booktabs}

\title{LVCNet: Efficient Condition-Dependent Modeling Network \\for Waveform Generation} 
\name{Zhen Zeng, Jianzong Wang\sthanks{Corresponding author: Jianzong Wang, jzwang@188.com}, Ning Cheng, Jing Xiao }
\address{Ping An Technology (Shenzhen) Co., Ltd.}
\begin{document}
%
\maketitle 
\begin{abstract}
    In this paper, we propose a novel conditional convolution network, 
    named location-variable convolution, to model the dependencies 
    of the waveform sequence. 
    Different from the use of unified convolution kernels in WaveNet to 
    capture the dependencies of arbitrary waveform, 
    the location-variable convolution uses convolution kernels 
    with different coefficients to 
    perform convolution operations on different waveform intervals, 
    where the coefficients of kernels is predicted 
    according to conditioning acoustic features, such as Mel-spectrograms.  
    Based on location-variable convolutions, we design LVCNet for waveform generation,  
    and apply it in Parallel WaveGAN to design more efficient vocoder.  
    Experiments on the LJSpeech dataset show that 
    our proposed model achieves a four-fold increase in synthesis speed compared to 
    the original Parallel WaveGAN without any 
    degradation in sound quality,  
    which verifies the effectiveness of location-variable convolutions. 

\end{abstract}
\section{Introduction}
\label{sec:intro}
In rencet years, deep generative models 
have witnessed extraordinary success in waveform generation, 
which promotes the development of speech synthesis systems 
with human-parity sounding.  
Early researches on autoregressive model in waveform synthesis, 
such as WaveNet \cite{wavenet} and WaveRNN \cite{WaveRNN}, 
have shown much superior performance over traditional 
parameters vocoders. 
However, low inference efficiency of autoregressive neural 
network limits its application in real-time scenarios.

In order to address the limitation and improve the generation speed, 
many non-autoregressive models have been studied to generate waveforms in parallel. 
One family relies on knowledge distillation, including 
Parallel WaveNet \cite{ParallelWavenet} and Clarinet \cite{Clarinet}, 
where an parallel feed-forward network is distilled from an autoregressive WaveNet model 
based on the inverse auto-regressive flows (IAF) \cite{IAF}. 
Although the IAF models is capable of generating high-fidelity speech in real time, 
the requirement for a well-trained teacher model and the intractable density distillation 
lead to a complicated model training process.  
The other family is flow-based generation models, including WaveGlow \cite{WaveGlow} 
and WaveFlow \cite{waveflow}. They are implemented by a invertible network 
and trained using only a single likelihood loss function on the training data. 
While inference is fast on high-performance GPU, the large size of mode limits their 
application in memory-constrained scenarios. 
Meanwhile, as a family of generation models, 
Generative Adversarial Network (GAN) \cite{GAN} is also applied in waveform generation, 
such as MelGAN \cite{MelGAN}, Parallel WaveGAN \cite{Parallel-WaveGan} and 
Multi-Band MelGAN \cite{multi-band-melgan}, 
in which a generator is designed to produce samples as close as possible to real speech, 
and a discriminator is implemented to distinguish generated speech from real speech. 
They have a very small amount of parameters, achieve a synthesis speech far exceeding real-time.
Impressively, the Multi-band MelGAN \cite{multi-band-melgan} runs at 
more than 10x faster than real-time on CPU. 
In addition, WaveGrad \cite{wavegrad} and DiffWave \cite{diffwave} 
apply diffusion probabilistic models \cite{diffusion} for waveform generation, 
which converts the white noise signal into structured waveform in an interative manner. 

These models are almost implemented by an wavenet-like network, 
in which the dilated causal convolution is applied to capture 
the long-term dependencies of waveform,  
and the mel-spectrum is used as the local conditional input 
for the gated activation unit. 
In order to efficiently capture time-dependent features, 
a large number of convolution kernels are required in wavenet-like network.
In this work, we propose the location-variable convolution 
to model time-dependent features more efficiently. 
In detail, the location-variable convolution uses convolution kernels 
with different coefficients to perform convolution operations 
on different waveform intervals, where the coefficients of kernels is predicted 
by a kernel predictor
according to conditioning acoustic features, such as mel-spectrograms.  
Based on location-variable convolutions, we design LVCNet for waveform generation,  
and apply it in Parallel WaveGAN to achieve more efficient vocoder.  
Experiments on the LJSpeech dataset \cite{ljspeech17} show that 
our proposed model achieves a four-fold increase in synthesis speed 
without any degradation in sound quality.
\footnote{Audio samples in \url{https://github.com/ZENGZHEN-TTS/LVCNet}} 

And the main contributions of our works as follow: 

\begin{itemize}
  \item A novel convolution method, named location-variable convolution, 
  is proposed to efficiently model the time-dependent features, 
  which uses different convolution kernels to perform convolution operations 
  on different waveform intervals;

  \item Based on location-variable convolutions, we design a network for waveform generation, 
  named LVCNet, and apply it in Parallel WaveGAN to achieve more efficient vocoder;

  \item A comparative experiment was conducted to demonstrate the effectiveness 
  of the location-variable convolutions in waveform generation.
\end{itemize}

\section{Proposed Methods}
\label{sec:proposed methods}

In order to model the long-term dependencies of waveforms more efficiently, 
we design a novel convolution network, named location-variable convolution, 
which is applied to the Parallel WaveGAN to verify its performance. 
The design details are described in the section. 

\subsection{Location-Variable Convolution}

In the traditional linear prediciton vocoder \cite{LinearPredictionVocoder1}, 
a simple all-pole linear filter is used to generate waveform in autoregressive way, 
of which the linear prediction coefficients is calculated according to the acoustic 
features. This process is similar to the autoregressive wavenet vocoder, except that 
the coefficients of the linear predictor is variable for different frames while 
the coefficients of convolution kernels in wavenet is the same in all frames. 
Inspired by this, we try to design a novel convolution network with variable convolution 
kernel coefficients in order to improve the ability to model 
long-term dependencies for waveform generation. 

\begin{figure}[t]
  \centering
  \includegraphics[width=0.7\linewidth]{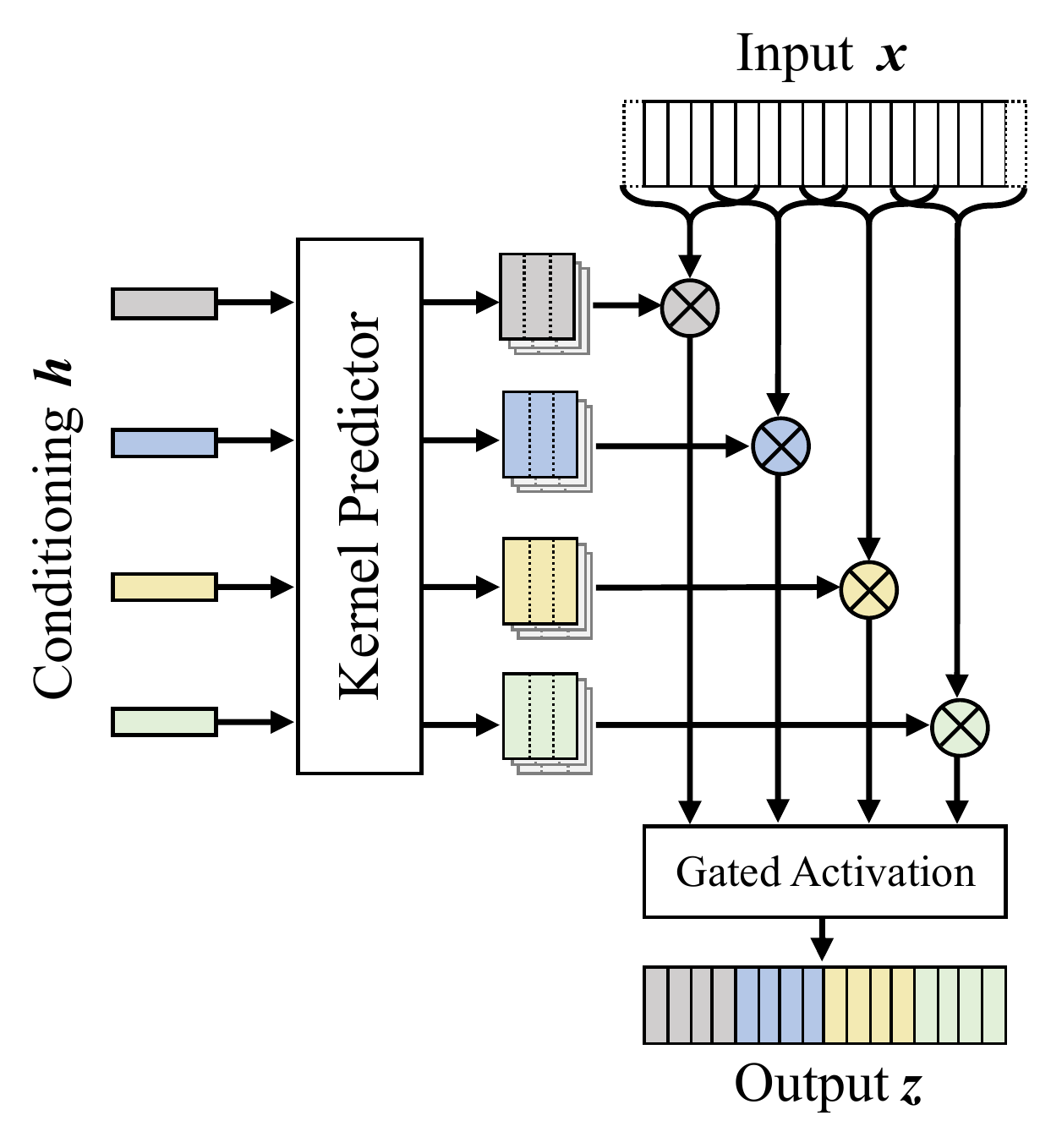}
  \caption{An example of convolution process in the location-variable convolution. 
  According to the conditioning sequence, the kernel predictor generates multiple sets of 
  convolution kernels, which are used to perform convolution operations on 
  the associated intervals in the input sequence. 
  Each element in the conditioning sequence corresponds to 4 elements in the input sequence. 
  }
  \label{figure-lvc}
\end{figure}

Define the input sequence to the convolution as 
$\boldsymbol{x}=\{x_1,x_2,$ $ \ldots, x_n\}$, 
and define the local conditioning sequence as 
$\boldsymbol{h}=\{h_1,h_2,\ldots,h_m\}$. 
An element in the local conditioning sequence is 
associated with a continuous interval in the input sequence. 
In order to effectively use the local correlation 
to model the feature of the input sequence,
the location-variable convolution uses a novel convolution method, 
where different intervals in the input sequence use 
different convolution kernels to implement the convolution operation. 
In detail, a kernel predictor is designed to predict multiple 
sets of convolution kernels according to the local conditioning sequence. 
Each element in the local conditioning sequence corresponds to a set 
of convolution kernels, which is used to perform convolution operations on 
the associated intervals in the input sequence. 
In other words, elements in different intervals of the input sequence 
use their related and corresponding convolution kernels to extract features. 
And the output sequence is spliced by the convolution results on 
each interval. 


Similar to WaveNet, the gated activation unit is also applied, 
and the local condition convolution can be expressed as 
\begin{align}
  & \{ \boldsymbol{x}_{(i)} \}_m = \text{split}( \boldsymbol{x} )  \\
  \{ \boldsymbol{W}^f_{(i)}, & \boldsymbol{W}^g_{(i)} \}_m = \text{Kernel Predictor}( \boldsymbol{h} ) \\
  \boldsymbol{z}_{(i)} &= \tanh ( \boldsymbol{W}^f_{(i)} * \boldsymbol{x}_{(i)} ) \odot \sigma ( \boldsymbol{W}^g_{(i)} * \boldsymbol{x}_{(i)} ) \\
  \boldsymbol{z} &= \text{concat} ( \boldsymbol{z}_{(i)} )
\end{align}
where $\boldsymbol{x}_{(i)}$ denotes the intervals of the input sequence associated with $ h_i $, 
$\boldsymbol{W}^f_{(i)}$ and $\boldsymbol{W}^g_{(i)}$ denote the filter and gate convolution kernels 
for $\boldsymbol{x}_{(i)}$. 

For a more visual explanation, Figure \ref{figure-lvc} shows an example of the location-variable convolution.
In our opinion, since location-variable convolutions can generate different kernels for different 
conditioning sequences, it has more powerful capability of modeling the long-term dependency than 
traditional convolutional network. 
We also experimentally analyze its performance for waveform generation in next section.

\begin{figure*}[t]
  \subfigure[ Architecture of LVCNet ]{
    \label{figure-lvcnet}
    \begin{minipage}[b]{0.4\linewidth} 
      \centering
      \includegraphics[width=0.55\linewidth]{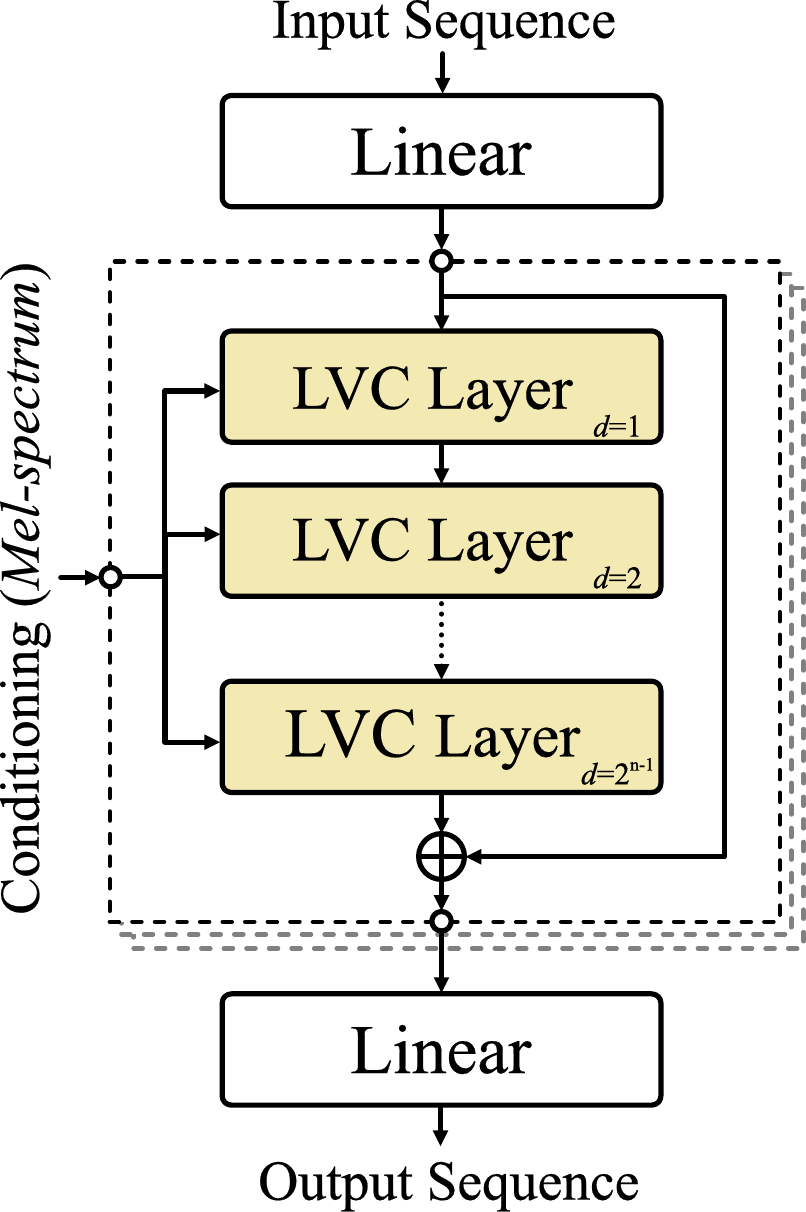}
    \end{minipage}
  }
  \subfigure[ Architecture of Parallel WaveGAN with LVCNet ]{
    \label{figure-lvcgan}
    \begin{minipage}[b]{0.59\linewidth} 
      \centering
      \includegraphics[width=0.75\linewidth]{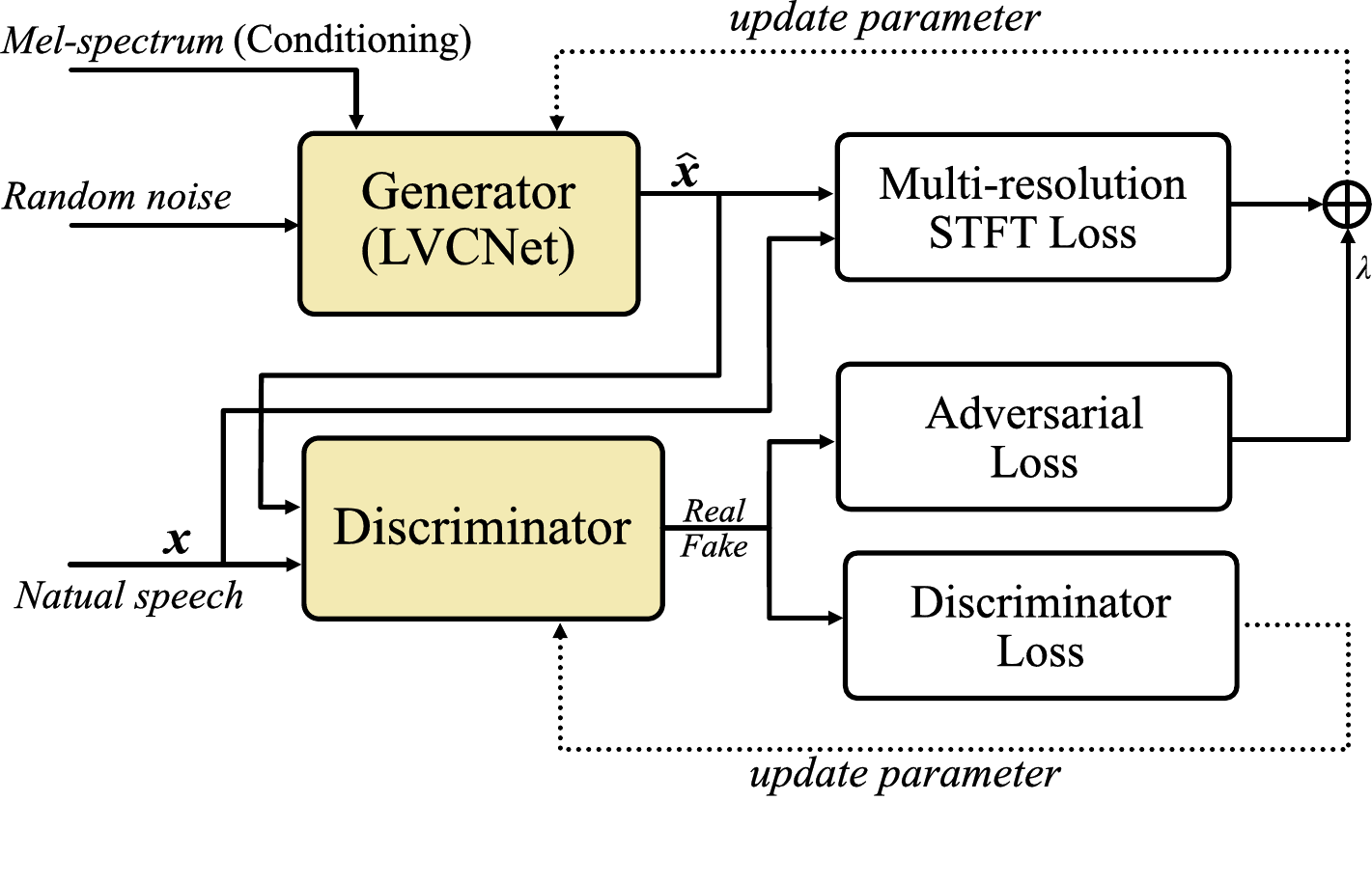}
    \end{minipage}
  }
  \caption{ (a) The architecture of LVCNet, which is composed of multiple LVCNet blocks, 
  and each LVCNet block contains multiple location-variable convolution (LVC) layers 
  with inscreasing factorial dilation coefficients to improve the receptive field.
  (b) The architecture of Parallel WaveGAN with LVCNet. The generator is implemented using LVCNet. }
\end{figure*}

\subsection{LVCNet} 

By stacking multiple layers of location-variable convolutions with different dilations, 
we design the LVCNet for waveform generation, as shown in Figure \ref{figure-lvcnet}. 
The LVCNet is composed of multiple LVCNet blocks, 
and each LVCNet block contains multiple location-variable convolution (LVC) layers 
with inscreasing factorial dilation coefficients to improve the receptive field.
A linear layer is applied on the input and output sides 
of the network to achieve channel conversion. 
The residual connection is deployed in each LVCNet block instead of each LVC layer, 
which can achieve more stable and satisfactory results according to our experimental analysis.
In addition, the conditioning sequence is input into the kernel predictor to
predict coefficients of convolution kernels in each LVC layer. 
The kernel predictor consists of multiple residual linear layer 
with leaky ReLU activation function ($\alpha = 0.1 $), 
of which the output channel is determined by the amount of convolution kernel 
coefficients to be predicted.

\subsection{Parallel WaveGAN with LVCNet}

In order to verify the performance of location-variable convolutions, 
we choose Parallel WaveGAN as the baseline model, 
and use the LVCNet to implement the network of the generator,  
as shown in Figure \ref{figure-lvcgan}. 
The generator is also conditioned on the mel-sepctrum, 
and transforms the input noise to the output waveform. 
For a fair comparison, 
the discriminator of our model maintains the same structure as that of Parallel WaveGAN, 
and the same loss function and training strategy as Parallel WaveGAN are used 
to train our model. 

Note that, the design of the kernel predictor is based on the correspondence between 
the mel-spectrum and the waveform. 
A 1D convolution with 5 of kernel size and zero of padding is firstly used to adjust the alignment 
between the conditioning mel-spectrum sequence and the input waveform sequence, 
and subsequent multiple stacked $1 \times 1$ convolutional layers to output 
convolution kernels of the location-variable convolutions.
In addition, we remove the residual connection 
of the first LVCNet block in the generator for better model performance.


\section{Experiments} 

\subsection{Experimental setup} 

\subsubsection{Database} 
We train our model on the LJSpeech dataset \cite{ljspeech17}. 
The entire dataset is randomly divided into 3 sets: 
12,600 utterances for training, 400 utterances for validation, and 100 utterances for test. 
The sample rate of audio is set to 22,050 Hz. 
The mel-spectrograms are computed through a short-time 
Fourier transform with Hann windowing, 
where 1024 for FFT size, 1024 for window size 
and 256 for hop size. The STFT magnitude is 
transformed to the mel scale using 80 channel 
mel filter bank spanning 80 Hz to 7.6 kHz. 

\subsubsection{Model Details}

In our proposed model, the generator is implemented by the LVCNet, 
and the network structure of the discriminator is consistent with 
that in original Parallel WaveGAN.  
The generator is composed of three LVCNet blocks, 
where each blocks contains 10 LVC layers, 
and the residual channels is set to 8. 
The kernel size of the location-variable convolution is set to three, 
and the dilation coefficients is the factorial of 2 in each LVCNet block.
The kernel predictor consists of one $1 \times 5 $ convolutional layer and 
three $1 \times 1$ residual convolutional layers, 
where the hidden residual channel is set to 64. 
The weight normalization is applied in all convolutional layer.

We choose Parallel WaveGAN \cite{Parallel-WaveGan} as the baseline model, 
and use the same strategy to train our model and baseline model 
for a fair comparison. 
In addition, in order to verify the efficiency of the location-variable convolution 
in modeling waveform dependencies, we conduct a set of detailed comparison experiments. 
Our proposed model is trained and compared with the Parallel WaveGAN 
under different residual channels 
(4, 8, 16 for our model and 32, 64, 128 for Parallel WaveGAN).

\subsection{Results}

\subsubsection{Evaluation} 
In order to evaluate the performance of these vocoder, 
we use the mel-spectrograms extracted from test utterances 
as input to obtain synthetic audios, which is rated together with 
the ground truth audio (GT) by 20 testers with headphones 
in a conventional mean opinion score (MOS) evaluation.
At the same time, the audios generated by Griffin-Lim 
algorithm \cite{GriffinLim} are also rated together. 

The scoring results of our proposed model and Parallel WaveGAN with different 
residual convolutional channels are shown in Table \ref{Table-Vocoder}, where 
the real-time factor implemented (RTF) on CPU is also illustrated.
We find that our proposed model achieves almost the same results as Parallel WaveGAN, 
but the inference speed is increased by four times. 
The reasan for the speech increase is that small number fo residual channels greatly reduces 
the amount of convolution operations in our model. 
Meanwhile, our unoptimized model can synthesizes multiple utterances at approximately 300 MHz 
on an NVIDIA V100 GPU, which is much faster than 35 MHz of Parallel WaveGAN. 

In addition, as the residual channels decreases, 
the rate of performance degradation of our model is 
significantly slower than that of Parallel WaveGAN. 
In our opinion, even if the residual channels is very small 
(such as 4, 8), the convolution coefficients are adjusted according to 
mel-spectrums in our model, which still guarantees effective feature modeling.

\subsubsection{Text-to-Speech} 

To verify the effectiveness of the proposed model as the vocoder in the TTS framework, 
we combine it with Transformer TTS \cite{TransformerTTS} and AlignTTS \cite{AlignTTS} for testing.  
In detail, according to the texts in the test dataset, Transformer TTS and AlignTTS predict 
mel-spectrums respectively, which is used as the input of our model (with 8 of residual channels) 
and Parallel WaveGAN to generate waveforms for MOS evaluation. 
The results are shown in Table \ref{Table-TTS}. 
Compared with Parallel WaveGAN, our model significantly improves the speed of speech synthesis 
without degradation of sound quality in feed-forward TTS systems.


In our opinion, due to the mutual independence of the acoustic features (such as mel-specturms), 
we can use difference convolution kernels to implement convolution operations 
on difference time intervals to obtain more effective feature modeling capabilities. 
In this work, we just use the LVCNet to design a new generator for waveform generation, 
and obtain a faster vocoder without degradation of sound quality. 
Considering our previous experiments \cite{MelGlow}, the effectiveness of 
the location-variable convolution has been sufficiently demonstrated, 
and there is potential for optimization. 

\begin{table}[t]
  \begin{center}
  \caption{The comparison between our proposed model (LVCNet) and Parallel WaveGAN (PWG) with different residual channels.}
  \begin{tabular}{p{1.7cm}p{1.2cm}<{\centering}p{1.8cm}<{\centering}p{1.9cm}<{\centering}}
  \toprule
  \textbf{Method}&\textbf{Size}&\textbf{MOS}&\textbf{ RTF (CPU) } \\
  \midrule
  GT           & & $ 4.56 \pm 0.05 $  &  $ - $ \\
  Griffin-Lim  & & $ 3.45 \pm 0.24 $  &  $ - $ \\
  PWG-32       & 0.44 M & $ 3.62 \pm 0.23 $  &  $ 2.16 $ \\   
  PWG-48       & 0.83 M & $ 4.03 \pm 0.10 $  &  $ 3.05 $ \\   
  PWG-64       & 1.35 M & $ 4.15 \pm 0.08 $  &  $ 3.58 $ \\   
  \midrule
  LVCNet-4        & 0.47 M & $ 3.96 \pm 0.15 $   &  $ 0.53 $ \\       
  LVCNet-6        & 0.84 M & $ 4.09 \pm 0.12 $   &  $ 0.62 $ \\       
  LVCNet-8        & 1.34 M & $ 4.15 \pm 0.10 $    &  $ 0.73 $ \\       
  \bottomrule
  \end{tabular}
  \label{Table-Vocoder}
  \end{center}
\end{table}

\begin{table}[t]
  \begin{center}
  \caption{The comparison between our proposed model (LVCNet) and Parall WaveGAN (PWG) in TTS systems.}
  \begin{tabular}{p{3.4cm}p{1.8cm}<{\centering}p{1.8cm}<{\centering}}
  \toprule
  \textbf{Method}&\textbf{MOS}&\textbf{Time (s)} \\
  \midrule
  GT & $-$ & $-$  \\
  Transformer + PWG   &   $ 4.08 \pm 0.14 $ &  $ 2.76 \pm 0.94 $  \\
  AlignTTS + PWG      &   $ 4.07 \pm 0.09 $ &  $ 0.09 \pm 0.01  $ \\
  \midrule
  Transformer + LVCNet   &   $ 4.07 \pm 0.15 $ &  $ 2.70 \pm 0.88  $ \\
  AlignTTS + LVCNet      &   $ 4.09 \pm 0.11 $ &  $ \bf{ 0.06 \pm 0.01 }  $ \\
  \bottomrule
  \end{tabular}
  \label{Table-TTS}
  \end{center}
\end{table}

\section{Conclusion}

In this work, we propose the location-variable convolution 
for time-dependent feature modeling. 
which uses different kernels to perform convlution operations 
on different intervals of input sequence.
Based on it, we design LVCNet and implement it as the generator 
of Parallel WaveGAN framework 
to achieve more efficient waveform generation model. 
Experiments on LJSpeech dataset show that our proposed model 
is four times faster than the base Parallel WaveGAN model in 
inferece speed without any degradation in sound quality, 
which verifies the effectiveness of the location-variable convolution.

\section{Acknowledgment}
This paper is supported by National Key Research and Development Program 
of China under grant No. 2017YFB1401202, No. 2018YFB1003500, and No. 2018YFB0204400. 
Corresponding author is Jianzong Wang from Ping An Technology (Shenzhen) Co., Ltd.

\bibliographystyle{IEEEbib}
\bibliography{reference}

\end{document}